\documentclass{article}
\usepackage{spconf,amsmath,graphicx}
\usepackage{etoolbox,siunitx}
\usepackage{multirow}
\usepackage{etoolbox,siunitx}
\robustify\bfseries
\usepackage{booktabs}
\usepackage{balance}
\usepackage{amssymb}
\usepackage{bm}
\usepackage{arydshln}
\usepackage{caption} 
\usepackage{xcolor}
\usepackage{enumitem, setspace, hyperref}
\usepackage{subcaption}

\input{def.set}

\DeclareMathAlphabet{\pazocal}{OMS}{zplm}{m}{n}
\newcommand{\Loss}{\pazocal{L}}

\title{Native Multi-Band Audio Coding\\within Hyper-Autoencoded Reconstruction Propagation Networks}

\name{Darius Petermann,$^{1}$
      Inseon Jang,$^{2}$
      Minje Kim,$^{1}$\thanks{This work was supported by Electronics and Telecommunications Research Institute (ETRI) grant funded by the Korean government (23ZH1200; ``The research of the basic media contents technologies").}}
\address{$^1$ Indiana University, Department of Intelligent Systems Engineering, Bloomington, IN  47408, USA\\
         $^2$ Electronics and Telecommunications Research Institute, Daejeon 34129, South Korea}

\begin{document}
\ninept
\maketitle
\begin{abstract}

Spectral sub-bands do not portray the same perceptual relevance. In audio coding, it is therefore desirable to have independent control over each of the constituent bands so that bitrate assignment and signal reconstruction can be achieved efficiently. In this work, we present a novel neural audio coding network that natively supports a multi-band coding paradigm. Our model extends the idea of compressed skip connections in the U-Net-based codec, allowing for independent control over both core and high band-specific reconstructions and bit allocation. Our system reconstructs the full-band signal mainly from the condensed core-band code, therefore exploiting and showcasing its bandwidth extension capabilities to its fullest. Meanwhile, the low-bitrate high-band code helps the high-band reconstruction similarly to MPEG audio codecs' spectral bandwidth replication. MUSHRA tests show that the proposed model not only improves the quality of the core band by explicitly assigning more bits to it but retains a good quality in the high-band as well.

\end{abstract}

\section{Introduction}
\label{sec:intro}

In recent years much focus has been put on the application of neural speech and audio coding specifically towards speech communication. %
While designing a neural codec depends substantially on the task at hand, recent research \cite{Zeghidour2021soundstream,jiang2022predictive} has opened venues for systems capable of coding both speech and non-speech audio signals (e.g., music), where these systems proposed to rely on generative power via adversarial loss for better perceptual quality as for audio coding. Another approach would be to introduce psychoacoustics to neural audio coding (NAC). Indeed, there has been a long history behind addressing the general audio compression problem from a perceptual standpoint through signal processing methods \cite{JohnstonJ1989midside, PainterT2000ieeeproc, DischS2016intelligentgap} and more recently perceptually-motivated loss functions for NAC \cite{KankanahalliS2018icassp, LiuQ2017perceptually, ZhenK2020spl, byun2022perceptual,defossez2022highfi}. 

Another less explored area in NAC is the multi-band consideration. For example, in speech coding, narrowing the signal bandwidth to 4~kHz or lower (i.e., a sampling rate of 8~kHz) has been a common approach \cite{hasegawa2003fundamentals}, where the \emph{core} of the intelligibility-related information usually lies in. 
While speech codecs are optimized mainly on the notion of intelligibility, such as found in neural speech codecs operating in very low bitrates (e.g., lower than 2.4kbps \cite{OordA2016wavenet, KleijnW2018wavenet, GarbaceaC2019vqvae}), the case for music tends to be different as the signals are sampled at a much higher rate, and consequently require a higher bitrate. Hence, codecs for this type of signal face different challenges as critical information is contained above the core band (CB). 

Simply focusing more on the high-frequency bands (HB) is not the most effective solution, as components in this area are perceptually less relevant. In general audio coding (e.g., music) tasks, there exists a trade-off between the effective audio bandwidth and the coding artifacts, which  codecs should be mindful of. For example, at a low bitrate (e.g., less than 128kbps for music), MPEG-2 Audio Layer III (a.k.a. MP3) \cite{mp3} would either start decreasing the effective bandwidth to save its effort in recovering less audible high frequency components: or else the lack of bits in the CB would introduce more audible coding artifacts. In order to partially circumvent this issue, low-bitrate modes of MPEG audio codecs, such as high-efficiency advanced audio coding (HE-AAC) \cite{HerreJ2008he-aac} or its MP3 version, MP3\emph{Pro}, introduced an effective alternative that exploits decreased perceptual sensitivity in the high-frequency area, namely ``spectral band replication" (SBR) \cite{sbr}. On the decoder side, SBR is based on the core-band reconstruction, from which the algorithm extracts patterns and replicates them at the higher end of the spectrum by using a considerably small amount of bits. 

From the perspective of machine learning, SBR could be interpreted as an informed version of bandwidth extension (BWE) or spectral recovery. BWE, in other applications, is usually defined as a ``blind" process that generates the HB component from CB with no additional information and is well-suited for machine learning \cite{SunD2013mlsp, EskimezS2019bandwidthex, gupta2019speechbe, su2021bandwidth}. In the context of NAC, BWE has been investigated as a method to recover the missing coefficients in the modified discrete cosine transform (MDCT) domain through a separate decoding process which uses quantized CB features \cite{kim2022recovery, shin2019recovery, shin2020enhanced}. 

Based on the success of both traditional SBR and neural BWE, we propose a multi-band approach to NAC. To this end, we re-design a U-Net-based NAC architecture, namely the hyper-autoencoded reconstruction propagation network (HARP-Net)~\cite{PetermannD2021harpnet}. It replaces the U-net's identity shortcut-based skip connections with small fully-convolutional autoencoders to compress information delivery between the encoder and the decoder. Our goal in this paper is to reflect a multi-band coding paradigm to HARP-Net, which we call MB-HARP-Net. In the simplest dual-band use case, we show that MB-HARP-Net provides a native interaction mechanism between the band-specific information paths for a better coding gain. MB-HARP-Net is conceptually similar to SBR and outperforms its dual-band coder baseline. %

MB-HARP-Net provides a principled neural mechanism for multi-band coding. It takes advantage of the intrinsic nature of strided convolution to perform \emph{neural downsampling}, which reduces a full-band input signal down to its CB component, on which neural compression and quantization are then performed to obtain a compact code. CB's code is then decoded back into its respective HB and CB components using two individual band-specific decoders, with the former path corresponding to \textit{neural upsampling} or bandwidth extension, while the latter is dedicated to CB reconstruction. In addition, the HB's decoder uses supplementary information coming through the autoencoded skip connection (i.e., a skip autoencoder (AE)). In contrast to \textit{blind} upsampling, the skip AE passes essential information for a more precise upsampling as in the SBR algorithm's small bitrate for the HB reconstruction. Hence, the harmonization of neural BWE and a low-bitrate code transfer via HARP-Net's skip AE is the key to a successful HB coding in the proposed architecture. 

Our experiments show that a frugal bit allocation to the HB reconstruction of MB-HARP-Net leads to an effective representation for both CB and HB, which optimize both bands' perceptual reconstruction quality. Although, for now, our proposed model falls short of the commercial HE-AAC v1 encoder's performance, it significantly outperforms other similar NAC architectures in the same bitrate thanks to the coupling of neural BWE and efficient HB coding. 
    
\section{Background}
\label{sec:background}

\subsection{Basic autoencoder}\label{sec:ae}

Given an input time-domain signal, $\bx\in\mathbb{R}^T$ (i.e., here understood as $\bx^{(0)}$), an AE typically performs audio compression by learning some hidden features $\bx^{(L)}$ from $\bx$ through an encoding function $\calF_{\text{enc}}$, e.g., a series of $L$ 1-d convolutional layers $f_\text{enc}^{(l)}$ ~\cite{KankanahalliS2018icassp}, where $l \in [1, \cdots, L]$: 
\begin{equation}
\bx^{(L)}=\calF_{\text{enc}}(\bx)=f_\text{enc}^{(L)} \circ f_\text{enc}^{(L-1)} \circ \cdots \circ f_\text{enc}^{(1)}(\bx) .
\end{equation}

A quantization function $Q$ can then be applied to the resulting ``deep'' encoded features $\bx^{(L)}$. $Q$ assigns the floating-point values from $\bx^{(L)}$ to a finite set of quantization bins (e.g., 32 centroids for a 5-bit system) to obtain the quantized feature vector $\overline{\bx}^{(L)}$. Hence, the transmission or storing can be done using the index to these $2^5$ centroids instead of the continuous feature vector ${\bx}^{(L)}$, reducing the bitrate. In addition, Huffman coding can introduce further bitrate reduction. On the receiver's end (i.e., the decoder), the code recovers the quantized feature vector as an estimate of  $\bx^{(L)} \approx \overline{\bx}^{(L)}$.

A decoding function $\calF_{\text{dec}}$ recovers the time-domain signal as $\overline{\bx}$ from $\overline{\bx}^{(L)}$, via another set of 1-d convolutional layers $f_\text{dec}^{(l)}, ~l \in [L, \cdots, 1]$, where $f_\text{dec}^{(1)}$ being the output layer. In a typical vanilla AE architecture, there is \textit{no} association between the encoder and decoder feature maps such that: $\bx^{(l-1)} \approx  f_\text{dec}^{(l)}(\overline{\bx}^{(l)})$.
In this paper, we rely on the U-Net's and, consequently, HARP-Net's assumption that this mirrored encoder-decoder reconstruction relationship should be enforced in the architecture for a better reconstruction. 

\subsection{Quantization and entropy}\label{sec:br}

\textbf{Quantization:}
The quantization process $Q$ considers the scalar feature assignment matrix $A^\text{hard}\in\mathbb{R}^{I\times J}$, which assigns one of the $J$ learned centroids $\bm{c}\in\mathbb{R}^J$ to each of the $I$ feature values in $\bz\in\mathbb{R}^I$. For example, in a vanilla AE, $\bz=\bm{x}^{(L)}$. During inference, the matrix $A^\text{hard}$ performs a non-differentiable ``hard" assignment, i.e., $\bz= A^\text{hard}\bm{c}$, where the $i$-th row of $A^\text{hard}$ is a one-hot vector selecting the closest centroid from the $i$-th feature. During training, in order to circumvent this non-differentiable process, we use a soft version of the assignment matrix $A^{\text{soft}}$, so the backpropagation error flows \cite{AgustssonE2017softmax}. The discrepancy between the soft and hard assignment results are handled by annealing a temperature factor $\alpha$ over the training iterations, so that $\overline{\bz} = A^{\text{hard}}\bm{c}=\text{lim}_{\alpha \rightarrow \inf} A^{\text{soft}}\bm{c}$. More formally, given a feature vector $\bz$ and centroids $\bm{c}$, we first create a distance matrix denoting the absolute difference between each element in $\bz$ and each centroid, $D\in\mathbb{R}^{I\times J}$. From $D$, we induce a probabilistic vector for each of the $i$-th element in $\bz$ as follows:
$A^{\text{soft}}_{i,:} = \text{softmax}(-\alpha D_{i,:})$.

\textbf{Bitrate control:}
We compute the empirical entropy of the quantized feature vector $\overline{\bz}$ by observing the assignment frequency of each of the centroids over multiple feature vectors. The assignment probability for centroid $j$ is therefore given by $p_j =P(c_j=A^\text{hard}_{i,:}\bc)$, which can be empirically approximated by $\overline{p}_j=
\frac{1}{IR}\sum_{i=1}^{IR} A^\text{hard}_{i,j}$ with $R$ denoting the total number of observable audio frames. Hence, the entropy estimate is given by: $\overline{H} =  - \ \sum_{j=1}^J \overline p_j \text{log}_2(\overline p_j)$.
To convert $\overline H$ into the lower bound of the bitrate counterpart $\overline B$, both $I$ (the code dimension) and the frame rate $F$ (i.e., the number of frames per second) need to be accounted for: $\overline B = F I \overline H$.

\subsection{HARP-Net}
\label{sec:harpnet}
In Sec. \ref{sec:ae}, the basic decoding process relies purely on the quantized feature vector $\overline{\bx}^{(L)}$ as its input, burdening the decoding function during reconstruction. HARP-Net \cite{PetermannD2021harpnet} repurposed the popular U-Net architecture \cite{RonnebergerO2015unet} for coding. It can circumvent the limitations of the basic AE with its mirrored U-Net architecture, whose pairs of corresponding encoder-decoder layers are interconnected via skip connections. What differentiates HARP-Net from U-Net is that each skip connection is replaced with a smaller-scale AE (i.e., ``skip'' AE) to compress and quantize the feature maps in delivery.

The skip AEs are denoted by $\calG^{(l)}$ with $l$ indicating the encoder-decoder layer pair. Since $\calG^{(l)}$ is also an AE, it is in its turn composed of an encoder $\calG_{\text{enc}}^{(l)}$ and decoder $\calG_{\text{dec}}^{(l)}$, as well as quantization module $Q$ which works as described in Sec. \ref{sec:br}. The feature maps from $f_\text{enc}^{(l)}$ can therefore be retrieved on the receiver side as follows: 
\begin{equation}
    \bx^{(l)}\!\approx\!\overline\bx_\calG^{(l)}\!\leftarrow\!\calG_\text{dec}^{(l)}(\overline{\bz}^{(l)}),~~~~
     \overline{\bz}^{(l)}\!\leftarrow\! Q (\bz^{(l)}), ~~~~
     \bz^{(l)}\!\leftarrow\!\calG_\text{enc}^{(l)}(\bx^{(l)}),\!\!
\end{equation}
where $\bz^{(l)}$ denotes the dimension-reduced version of the intermediate feature maps at the $l$-th layer pair $\bx^{(l)}$. The $f_\text{dec}^{(l)}$ then takes the concatenation of the skip AE's output to the output of its preceding layer $(l+1)$ to produce its output: $\overline{\bx}^{(l-1)}\leftarrow f_\text{dec}^{(l)}\big([\overline{\bx}^{(l)}, \overline{\bx}_\calG^{(l)}]\big)$.

The additional skip AEs in HARP-Net create information paths that circumvent the aggressive bottleneck-only quantization process at the $L$-th layer. This comes at the cost of a more sophisticated bitrate computation by aggregating all bitrates computed at up to $L$ different codes, which we denote by $\overline B=\sum_{l=1}^L FI^{(l)}\overline H^{(l)}$. The bit allocation strategy among different skip AEs has potential, but it was not fully studied in the original HARP-Net paper.

\section{The Proposed Multi-Band HARP-Net}

\begin{figure}
    \centering
        \includegraphics[width=1.0\linewidth]{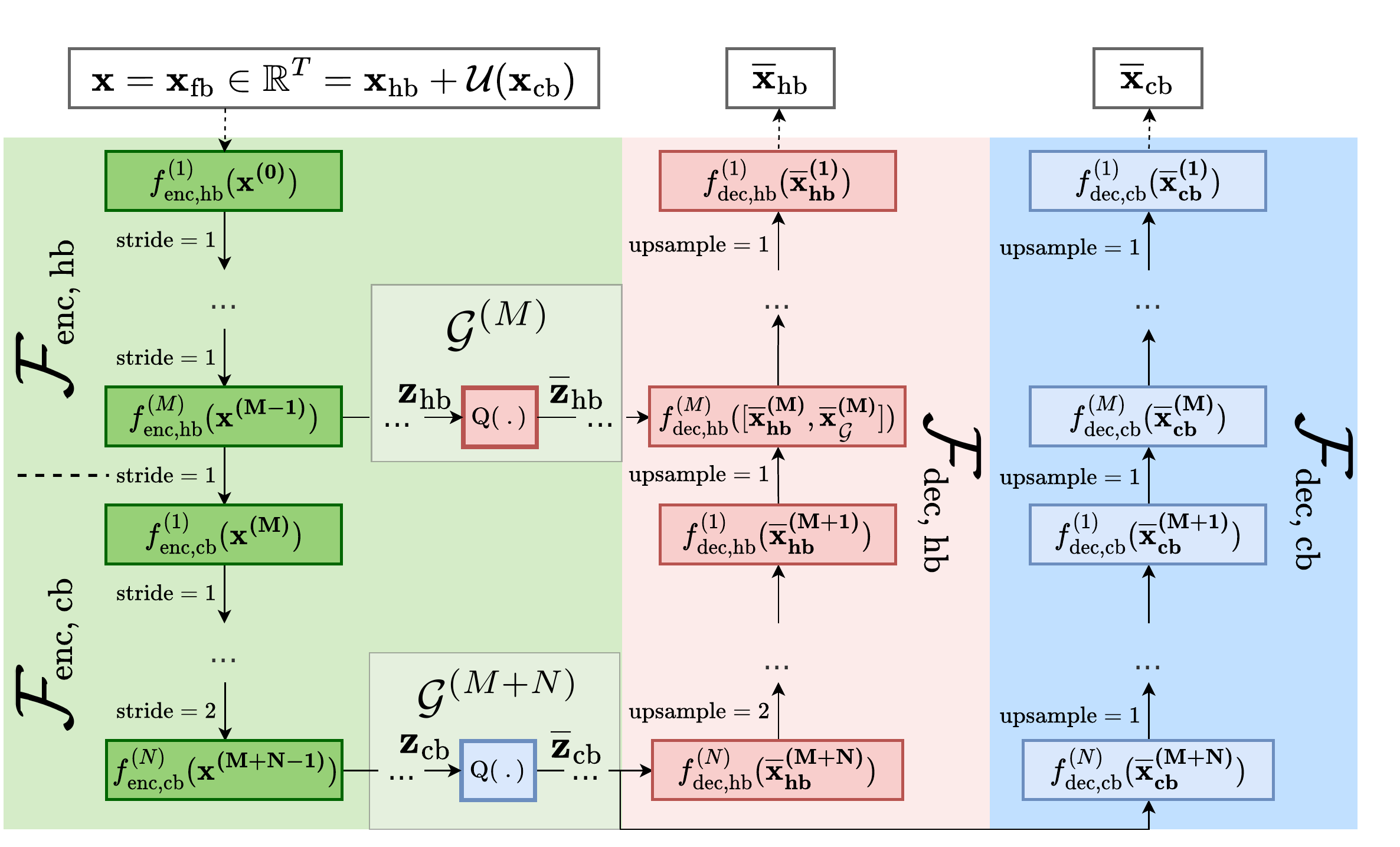}
    \caption{Overview of our proposed MB-HARP-Net architecture.}
    \label{fig:overall_example}
\end{figure}

\subsection{Cascaded encoders architecture}

While HARP-Net has proven to help the total AE reconstruction compared to a vanilla AE, it is up to the optimization process as to how the target bitrate is fragmented and assigned among all $\bz$'s $[\bz^{(L)}; \bz^{(L-1)}; \cdots \bz^{(1)}]$.  That is, there is no constraint as to what type of feature representations are learned through all the feature vectors $\bz^{(l)}$'s. For example, the deepest feature, $\bz^{(L)}$, would better get assigned the most amount of bits as they represent the most abstract and structural information of the original input $\bx$. 

In this work, we aim to harmonize BWE and neural audio coding; therefore, explicitly enforcing sub-band (i.e., CB and HB) representations is the first required step. MB-HARP-Net is based on the HARP-Net's convenient architectural paradigm which already provides various abstract levels of feature representations, while they are not related to multi-band coding yet. MB-HARP-Net naturally extends the idea of skip AEs, allowing for independent control over both band-specific reconstructions and bitrate assignments. 

More formally, we redefine our full-band input time-domain signal $\bx$ as the sum of the two sub-band signals $\bx=\bx_{\text{fb}}\in\mathbb{R}^{T} = \bx_\text{hb} + \mathcal{U}(\bx_\text{cb})$, where $\bx_\text{hb}$ and $\bx_\text{cb}$ stand for high-pass filtered and downsampled versions of the full-band signal, respectively.  As for the downsampled one, we make sure the decimated signal $\bx_\text{cb}$ recovers the original temporal resolution via an interpolation-based upsampling $\mathcal{U}$ during the inference. During training however, the reconstruction loss is computed using the downsampled version $\bx_\text{cb}$. 

The encoder $\calF_{\text{enc}}$ is now split into two \emph{cascaded} stages; $\calF_{\text{enc, hb}}$ consists of $M$ 1-d convolutional layers and takes in the full-band input $\bx$. Since its stride is one and due to zero padding, the original input temporal dimension $T$ remains intact. In contrast, the channel dimension increases to $C$, which is the number of filters. 
The second encoding stage, $\calF_{\text{enc, cb}}$, takes $\bx^{(M)}$ as input. Some of its $N$ 1-d convolutional layers have $\text{stride}>1$, resulting in a decimating factor of $\delta^{\text{ds}}=\prod_k \delta^{\text{ds}}_k$, where $\delta^{\text{ds}}_k$ is the stride of the participating layer $k$. As oppposed to  $\bx^{(M)}$, the output of $f^{(N)}_\text{enc, cb}$, $\bx^{(M+N)}$, loses the high-frequency content during the temporal decimation $T/\delta^{\text{ds}}$. However, its temporal structure that corresponds to $\bx_\text{cb}$ remains unaffected. 

\subsection{Band-wise quantization procedure}

The output of both encoding stages $\bx^{(M)}$ and $\bx^{(M+N)}$, respectively, are fed to their own dedicated skip AEs, $\calG_\text{enc}^{(M)}$ and $\calG_\text{enc}^{(M+N)}$ through which channel reduction and quantization are performed. Each skip AE's encoder does not perform any downsampling on its input features. Instead, the channel dimension collapses from $C$ down to 1 at its own bottleneck layer: $\bz_\text{hb}\in\mathbb{R}^{T\times 1}\leftarrow\calG_\text{enc}^{(M)}(\bx^{(M)})$ and 
$\bz_\text{cb}\in\mathbb{R}^{T/\delta\times 1}\leftarrow\calG_\text{enc}^{(M+N)}(\bx^{(M+N)})$. This channel reduction decreases the data rate while the temporal resolution is maintained. 

As described in Sec. \ref{sec:br}, each of the code vector values is assigned one of the $J$ centroids which are learned independently for each of the sub-bands; $\bm{c}_\text{cb}$ and $\bm{c}_\text{hb}$. We assume the code distributions (i.e., the numbers of learned centroids $J_\text{hb}$ and $J_\text{cb}$, and their bitrates) $B_\text{cb}$ and $B_\text{hb}$ to be learned and controlled individually.

\subsection{Multi-band decoder architecture}
The decoding process of the MB-HARP-Net respects the multi-band nature of the code structure by dedicating a decoding head to each band. First, the quantized code vectors $\overline{\bz}_{\text{cb}}$ and $\overline{\bz}_{\text{hb}}$ are decoded back into their corresponding tensor representation $\overline{\bx}^{(M)}$ and $\overline{\bx}^{(M+N)}$ using their respective skip AE's decoding stage:
\begin{equation}
\begin{aligned}
    \bx^{(M)} \approx \overline{\bx}^{(M)}_\calG & \leftarrow \calG^{(M)}_{\text{dec, hb}}\circ Q(\bz_\text{hb}) \\ 
    \bx^{(M+N)} \approx \overline{\bx}^{(M+N)}_\calG & \leftarrow \calG^{(M+N)}_{\text{dec, cb}}\circ Q(\bz_\text{cb})
\end{aligned}
\end{equation}
which are then used as the input to the two band-specific HARP-Net decoding heads.
Both band-specific decoding paths $\mathcal{F}_\text{dec, cb}$ and $\mathcal{F}_\text{dec, hb}$ take $\overline{\bm x}^{(M+N)}_\calG$ as their direct input, while $\mathcal{F}_\text{dec, hb}$ takes additional input from the coded skip connection $\overline{\bm x}^{(M)}_\calG$ for HB reconstruction at the $M$-th layer.  

\textbf{CB reconstruction}: The CB decoding stack $\calF_{\text{dec, cb}}$ shares the same depth as the encoding counterpart $(M+N)$ and recovers the $T/\delta^{\text{ds}}$-dimensional downsampled version of the input signal as follows: ${\bx}_{\text{cb}} \approx \overline{\bx}_{\text{cb}}=\calF_{\text{dec, cb}}(\overline{\bx}^{(M+N)}_\calG)$.

\textbf{HB reconstruction}: In parallel, the HB reconstruction is conducted via a separate decoding path ${\bx}_{\text{hb}} \approx \overline{\bx}_{\text{hb}} = \calF_{\text{dec, hb}}(\overline{\bx}^{(M+N)}_\calG)$. Note here that while no explicit high-pass filtering is applied during the HB reconstruction path, it is implicitly enforced as the reconstruction is compared to the high-pass filtered signal $\bx_\text{hb}$. The primary difference between $\calF_{\text{dec, hb}}$ and $\calF_{\text{dec, cb}}$ is that the former specifically employs nearest-neighbors-based \textit{upsampling} operations during its CNN-based operations \cite{pons2021upsampling} to make up the loss of temporal resolution. Hence, $\calF_{\text{dec, hb}}$ can be seen as a neural BWE process whose performance would not be perfect. Upsampling will increase the temporal resolution of the final signal by $\delta^{\text{us}}=\prod_k \delta_k^{\text{us}}$, where $\delta^{\text{us}}_k$ is the scaling factor of the participating layer $k$.

In addition to the BWE-corresponding decoding operation, $\calF_{\text{dec, hb}}$ leverages the role of $\overline{\bm z}_{\text{hb}}$ (i.e., the HB-only code). After the skip AE's decoding stage, $\overline{\bm x}^{(M)}_\calG$ is concatenated with $\overline{\bm{x}}_\text{hb}^{(M)}$ along the channel dimension, forming the following input at the $M$-th layer of $\calF_{\text{dec, hb}}$
\begin{equation}
\label{eq:hb_concat}
    \overline\bx_\text{hb}^{(M-1)} \leftarrow f_\text{dec,hb}^{(M)}\left(\big[\overline{\bm x}_\text{hb}^{(M)}, \overline{\bm x}^{(M)}_\calG\big]\right).
\end{equation}
Here, $f_\text{dec, hb}^{(M)}$ takes  $2C$ channels and collapses them down to $C$ channels. Note that we distinguish $\overline{\bm x}_\text{hb}^{(l)}$ from $\overline{\bm x}_\text{cb}^{(l)}$ as there are two decoding heads generating respective feature maps concurrently. 

\subsection{Band-wise entropy control and objective function}

We design the objective function so the codec provides the best reconstruction while reaching the target entropy of both codes $\overline{\bz}_{\text{cb}}$ and $\overline{\bz}_{\text{hb}}$. The network loss consists of reconstruction and entropy control terms. The reconstruction error is measured via both time and frequency domain losses: negative signal-to-noise ratio (SNR) and $L_1$ norm over the log magnitudes of short-time Fourier transform (STFT). In addition, each is with two sub-band specific versions. We denote this reconstruction loss as a weighted sum: $\mathcal{L}_\text{recons}=\sum_{b\in\{\text{cb, hb}\}}\sum_{d\in\{\text{SNR, STFT}\}}\lambda_{b,d}\mathcal{L}_{b,d}$ with the blending weights $\lambda_{b,d}$.

\begin{figure*}[t]
    \centering
        \begin{subfigure}[b]{0.222\textwidth}
         \centering
         \includegraphics[height=1.475cm]{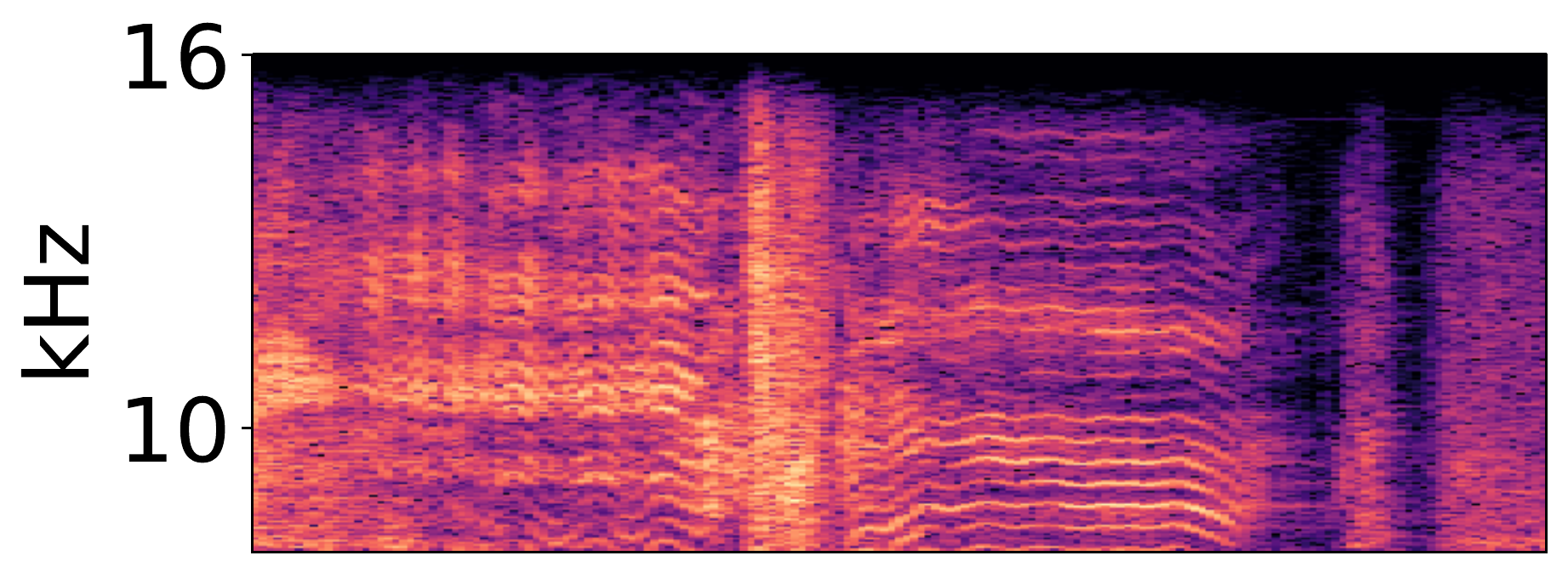}\vspace{-0.2cm}
         \caption{$\bm{x}_\text{hb}$}
     \end{subfigure}
     \hfill
     \begin{subfigure}[b]{0.19\textwidth}
         \centering
         \includegraphics[height=1.39cm]{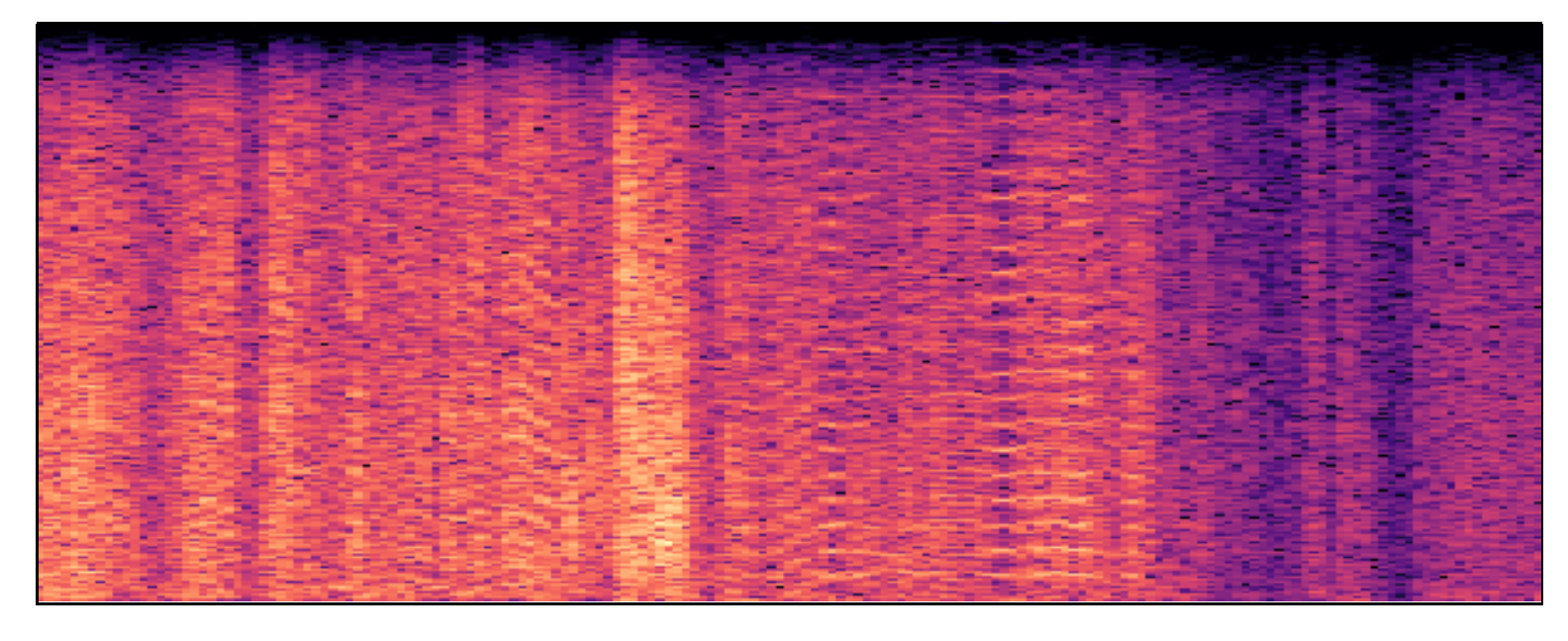}\vspace{-0.2cm}
         \caption{\texttt{bl1}: $\overline{\bm{x}}_\text{hb}; [\bm{z}_\text{cb}, \bm{z}_\text{hb}]$}
     \end{subfigure}
     \hfill
     \begin{subfigure}[b]{0.19\textwidth}
         \centering
         \includegraphics[height=1.39cm]{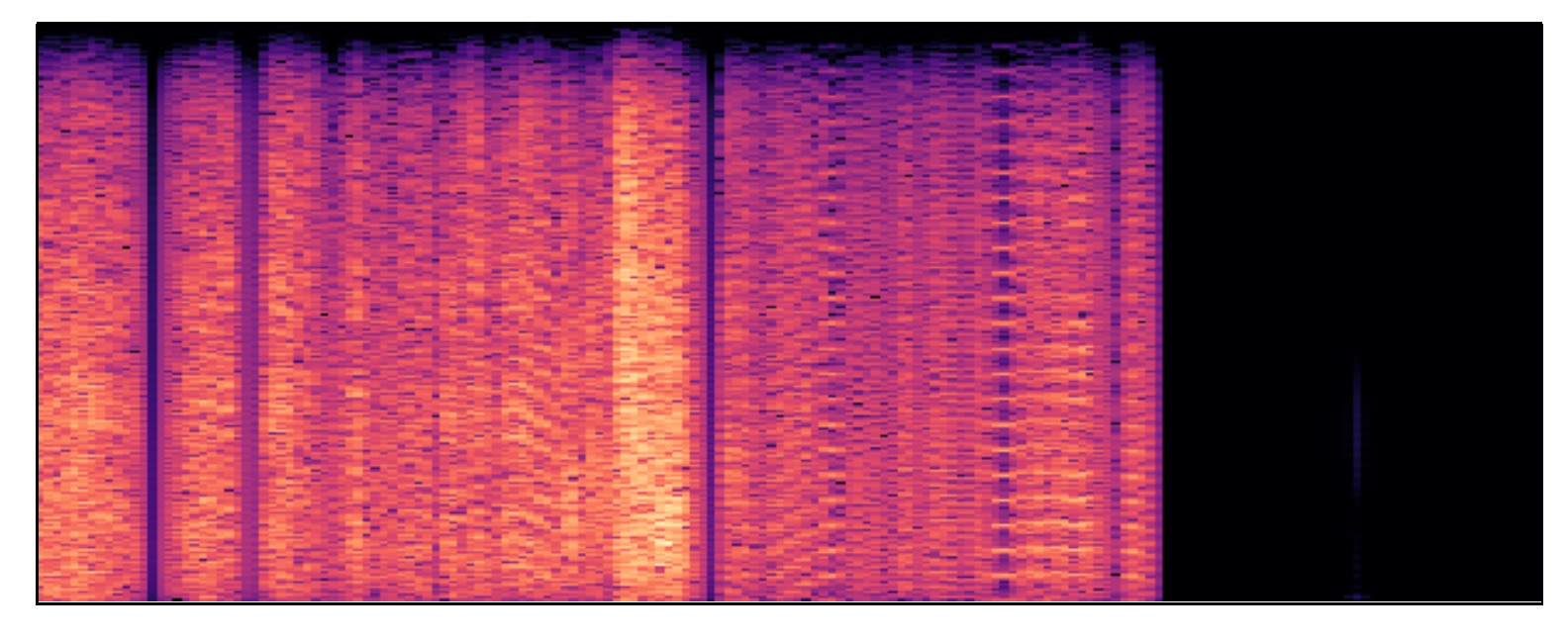}\vspace{-0.2cm}
         \caption{\texttt{bl1}: $\overline{\bm{x}}_\text{hb}; [\bm{0}, \bm{z}_\text{hb}]$}
     \end{subfigure}
     \begin{subfigure}[b]{0.19\textwidth}
         \centering
         \includegraphics[height=1.39cm]{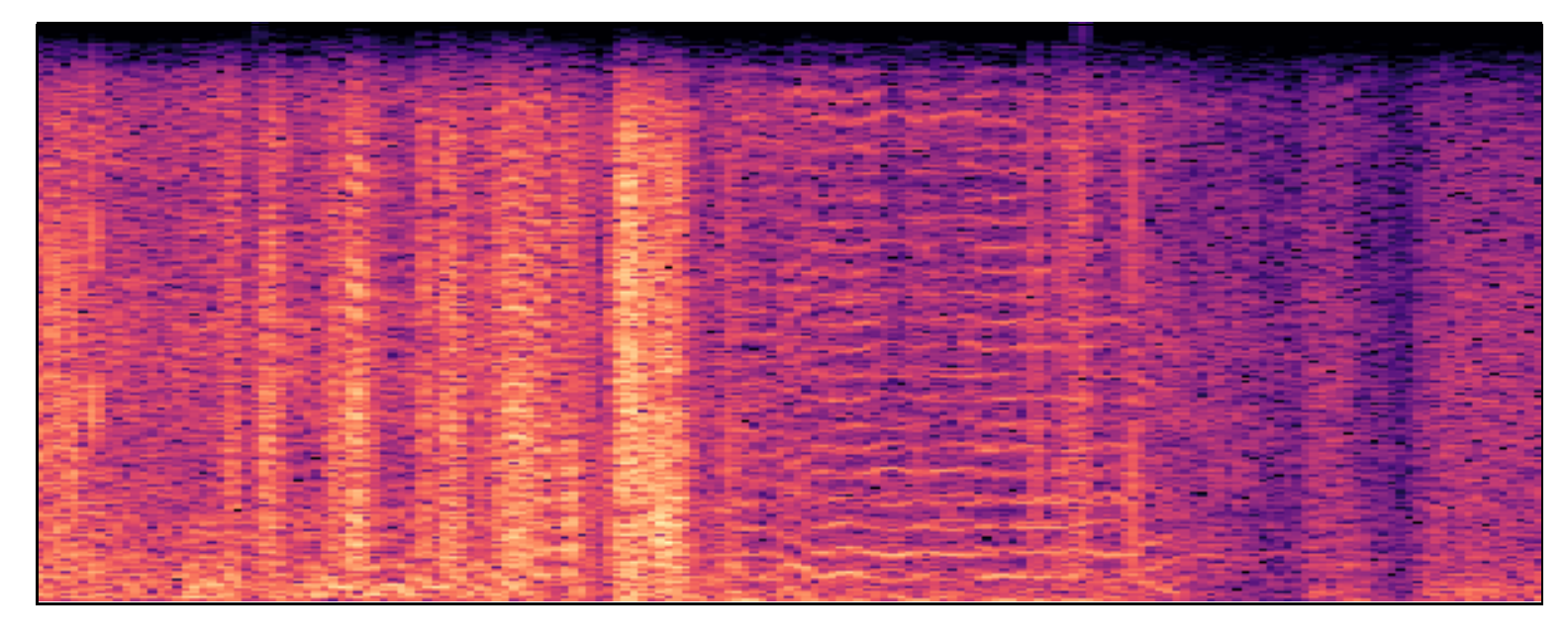}\vspace{-0.2cm}
         \caption{\texttt{p}$_{B_{34}:B_6}$: $\overline{\bm{x}}_\text{hb}; [\bm{z}_\text{cb}, \bm{z}_\text{hb}]$}
     \end{subfigure}
     \hfill
     \begin{subfigure}[b]{0.19\textwidth}
         \centering
         \includegraphics[height=1.39cm]{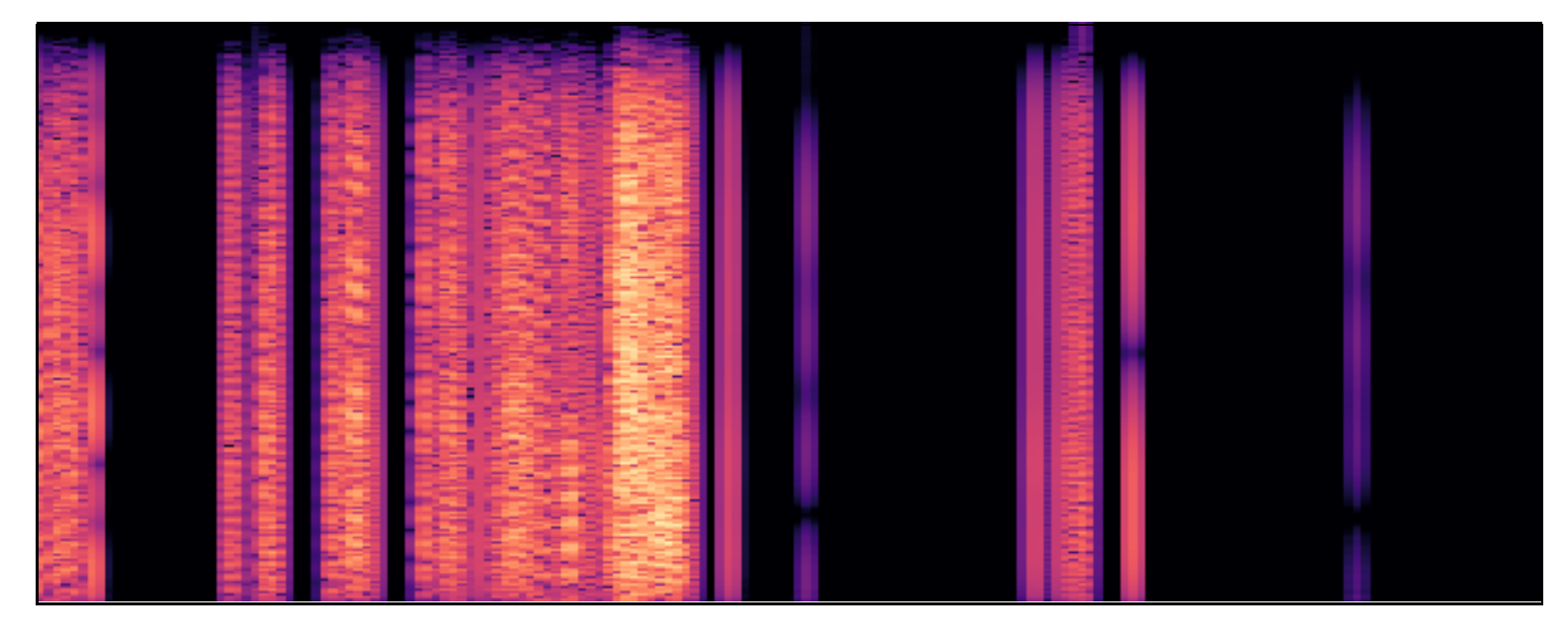}\vspace{-0.2cm}
         \caption{\texttt{p}$_{B_{34}:B_6}$: $\overline{\bm{x}}_\text{hb}; [\bm{0}, \bm{z}_\text{hb}]$}
     \end{subfigure}
    \vspace*{-3mm}
    \caption{The impact of missing CB code in HB reconstruction. $[\bm{0}, \bm{z}_\text{hb}]$ is the case with no CB code, while $[\bm{z}_\text{cb}, \bm{z}_\text{hb}]$ is the regular case.}
    \label{fig:ablation_spectrum}
\end{figure*}

\begin{figure}[t]
    \centering
        \includegraphics[width=1.0\linewidth]{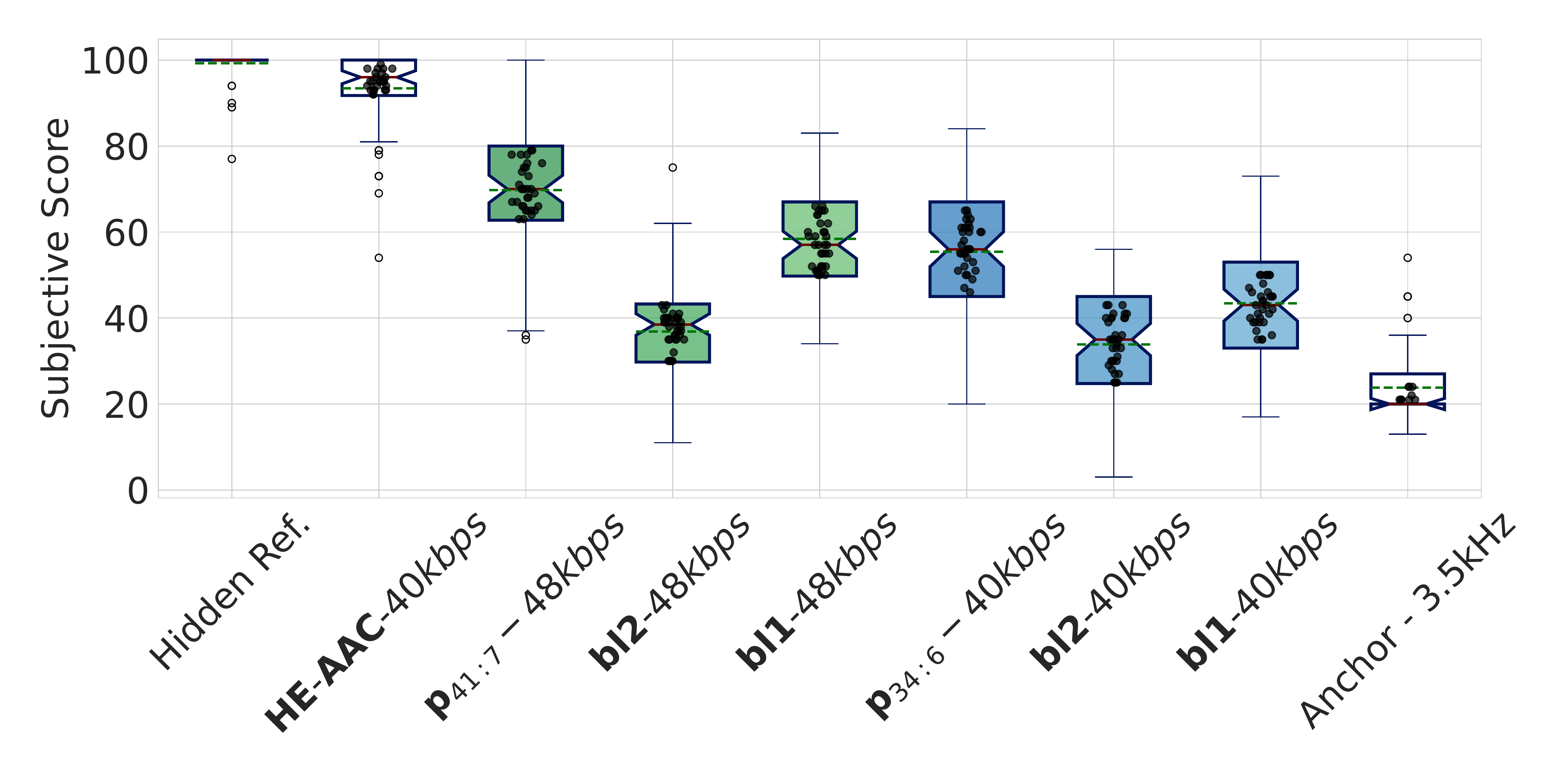}\vspace{-0.5cm}
    \caption{MUSHRA test results; green dotted lines depict the mean while the red line represents the median. The proposed model's dual-band bitrate assignments are represented as $\texttt{p}_{{B_\text{cb}}:B_{\text{hb}}}$. }
    \label{fig:mushra_results}
\end{figure}

The bitrates of the system is regularized by band-specific entropy loss terms, which can be represented in terms of bitrates as described in \ref{sec:br} and \ref{sec:harpnet}:
$\Loss_{\text{br}} = \lambda_{\text{br}} \ \sum_{b\in\{\text{cb, hb}\}} |B_b - \overline B_b|$,
where $\overline B_b$ denotes the bitrate estimated from the band-specific code vector with the index $b$.
$B_b$ is the manually defined band-specific target bitrate. $\lambda_{\text{br}}$ controls the contribution of the entropy loss to the total loss which is defined as $\Loss = \Loss_{\text{br}}+\mathcal{L}_\text{recons}$.

\section{Experimentals}
\label{sec:experiment}

Our experiments aim at two bitrates, 40kbps and 48kbps, with a main aim to showcase the benefits of MB-HARP-Net over its single-band counterpart. The proposed model is based on frame-by-frame processing, with each input frame being $T=16,384$ samples (i.e., $\approx 0.5$s) and with an overlap size of $32$ samples that are windowed using a Hann function. Commercial music signals are used in the experiments; they entail $6,000$ song segments across 13 genres, totaling in 5.5 hours of audio data that are partitioned into 80\%, 15\%, and 5\% for training, validation, and testing sets, respectively. All the signals are monophonic and natively sampled at 44.1~kHz, although we downsample them to 32~kHz to match the HE-AAC specifications \cite{HerreJ2008he-aac} for the target bitrates. 

Adam optimizer \cite{KingmaD2015adam} is used with validation loss-based early stopping. The training process also monitors the entropy loss $\mathcal{L}_\text{br}$ so that it conforms to our target bitrate region ($\pm 1.5$ kbps). $M$ and $N$ are both set to be 3, i.e., the total depth is $M+N=6$. Downsampling is introduced at the last layer of $\calF_{\text{enc, cb}}$, i.e.,$\delta^{\text{ds}}_{k=N} = 2$. Likewise, in $\calF_{\text{dec, hb}}$, upsampling is done in the first layer: $\delta^{\text{us}}_{k=M+N} = 2$. Each layer consists of $C=50$ 1-d convolutional kernels of size $15$. Following the Fraunhofer's open-sourced implementation\footnote{https://github.com/mstorsjo/fdk-aac}, we roughly match the cutoff frequencies: 8~kHz for $\bx_\text{cb}$ and 7.3~kHz for $\bx_\text{hb}$, where the overlapping band of 700Hz is to make up for the filtering loss of energy around the cutoffs. $\lambda_{b,\text{STFT}}$ and $\lambda_{\text{br}}$ are set to 15.0 (for both bands) and $6 \times 10^{-4}$, respectively. $\lambda_{b, \text{SNR}}$, however, varies depending on the model trained (see following descriptions).

\textbf{Models:} To assess the impact and benefit of band-specific entropy control, we propose to assess three models that share the same architecture but are regularized differently.
\begin{itemize}[noitemsep,topsep=0pt, leftmargin=0in, itemindent=.15in]
    \item \texttt{bl1}: No explicit band-wise bitrate assignment is applied; rather, we let the optimization process decides bit allocation (e.g., as we do in HARP-Net). $\lambda_{b,\text{SNR}}=1$ for both bands. In this model, the reconstruction loss tends to promote CB reconstruction rather than HB as they lack modeling power.
    \item \texttt{bl2}: 
    \texttt{bl2} scales up $\lambda_{\text{hb,SNR}}$ to 2.0 while maintaining $\lambda_{\text{cb,SNR}}=1.0$ for better HB reconstruction. \texttt{bl2} controls the sub-band reconstructions individually via optimization, while its bitrate control does not fully benefit from the MB-HARP-Net architecture yet. 
    \item \texttt{p}$_{B_\text{cb}:B_\text{hb}}$: Our proposed model enforces band-specific bitrate assignments: much more bits to CB than to the HB, but with twice larger blending weight $\lambda_{\text{hb,SNR}}=2$ as in \texttt{bl2}. ${B_\text{cb}}:B_{\text{hb}}$ defines the actual bit allocation ratio that the model is trained to target on. Differently from a HARP-Net version's per-skip-AE bitrate control, the proposed model benefits from (a) the upsampling path's blind BWE (b) the supplementary HB code $\overline{\bm{z}}_\text{hb}$, similarly to what SBR does.
\end{itemize}

\textbf{Subjective Listening Tests:}
Ten audio experts participated in our MUSHRA test \cite{mushra} on 8 different test songs. Each trial includes our three systems in comparison at both bitrates (40 and 48kbps), HE-AAC v1 encoded at 40kbps, a hidden reference, and a low-pass filtered anchor at 3.5~kHz. The items were selected to represent the diversity of 16 music genres used in our proprietary music dataset. Subjects used high-quality headphones. Fig.~\ref{fig:mushra_results} shows the subjective test results. We observe that our proposed systems \texttt{p}$_{B_{\text{cb}}:B_{\text{hb}}}$ outperform both of their baseline counterparts in the two bitrate cases. Our 40kbps model \texttt{p}$_{{\text{34}}:{\text{6}}}$ competes with \texttt{bl1} at 48kbps, showcasing the proposed model's superiority in reconstructing HB components while consuming fewer bits. Meanwhile, a deliberate effort to focus more on the HB reconstruction, as in \texttt{bl2}, fails to achieve the desired perceptual quality because it is merely based on the weighting of HB reconstruction loss, leading to a poor reconstruction in the perceptually important CB area. The proposed methods underperform the commercial HE-AAC with SBR (on Adobe$^\text{\textregistered}$ Audition$^\text{\textregistered}$). 

\textbf{Ablation Study:}
To explore MB-HARP-Net's BWE capabilities, we perform an ablation experiment where $\overline{\bz}_{\text{cb}}$ is either included as part of $\overline{\bx}_\text{hb}$ reconstruction or not. This allows us to explore the relevance of $\overline{\bz}_{\text{cb}}$'s contribution towards $\overline{\bx}_\text{hb}$. Fig. \ref{fig:ablation_spectrum} illustrates the impact of excluding $\bm{z}_\text{cb}$ from the HB reconstruction. We observe that \texttt{bl1} still shows reasonable HB reconstruction without a proper neural upsampling-based BWE, demonstrating its significant reliance on the HB code 
$\bm{z}_\text{hb}$. It is not a desirable usage of bits given that the HB area is perceptually less relevant. On the other hand, our \texttt{p}$_{B_{34}:B_{6}}$ shows poorer HB reconstruction quality if it were not for $\bm{z}_\text{cb}$. That is, the proposed model's coding gain comes from the bitrate-free BWE process, while saving bits on $\bm{z}_\text{hb}$. Note that we decided to exclude the favorable results from the SNR-based objective measures, as they can also result in misleading conclusions in audio coding especially in the context of band-specific reconstructions.

\section{Conclusions}
\label{sec:conclusion}

In this work we presented a multi-band paradigm to re-designing a U-Net-based neural audio coding system, namely HARP-Net, that we call MB-HARP-Net. Our algorithm offers a neural mechanism to enforce band-specific representations, which allows for independent control over the sub-bands' reconstruction as well their respective bitrate assignment. Since both core and high-bands are principally reconstructed from the core-band code vector, MB-HARP-Net draws a neural analog to SBR, which couples neural bandwidth extension and efficient sub-band coding. We opensourced the project at: \url{https://saige.sice.indiana.edu/research-projects/HARP-Net}.

\newpage
\bibliographystyle{IEEEtran}
\bibliography{mjkim}

\begin{thebibliography}{10}
\providecommand{\url}[1]{#1}
\csname url@samestyle\endcsname
\providecommand{\newblock}{\relax}
\providecommand{\bibinfo}[2]{#2}
\providecommand{\BIBentrySTDinterwordspacing}{\spaceskip=0pt\relax}
\providecommand{\BIBentryALTinterwordstretchfactor}{4}
\providecommand{\BIBentryALTinterwordspacing}{\spaceskip=\fontdimen2\font plus
\BIBentryALTinterwordstretchfactor\fontdimen3\font minus
  \fontdimen4\font\relax}
\providecommand{\BIBforeignlanguage}[2]{{%
\expandafter\ifx\csname l@#1\endcsname\relax
\typeout{** WARNING: IEEEtran.bst: No hyphenation pattern has been}%
\typeout{** loaded for the language `#1'. Using the pattern for}%
\typeout{** the default language instead.}%
\else
\language=\csname l@#1\endcsname
\fi
#2}}
\providecommand{\BIBdecl}{\relax}
\BIBdecl

\bibitem{Zeghidour2021soundstream}
N.~Zeghidour, A.~Luebs, A.~Omran, J.~Skoglund, and M.~Tagliasacchi,
  ``Soundstream: An end-to-end neural audio codec,'' \emph{IEEE/ACM Trans.
  Audio, Speech and Lang. Proc.}, vol.~30, p. 495–507, jan 2022.

\bibitem{jiang2022predictive}
X.~Jiang, X.~Peng, H.~Xue, Y.~Zhang, and Y.~Lu, ``Predictive neural speech
  coding,'' \emph{arXiv preprint arXiv:2207.08363}, 2022.

\bibitem{JohnstonJ1989midside}
J.~D. {Johnston}, ``Perceptual transform coding of wideband stereo signals,''
  in \emph{Proc. of the IEEE International Conference on Acoustics, Speech, and
  Signal Processing (ICASSP)}, 1989, pp. 1993--1996 vol.3.

\bibitem{PainterT2000ieeeproc}
T.~Painter and A.~Spanias, ``Perceptual coding of digital audio,''
  \emph{Proceedings of the IEEE}, vol.~88, no.~4, pp. 451--515, 2000.

\bibitem{DischS2016intelligentgap}
S.~Disch \emph{et~al.}, ``Intelligent gap filling in perceptual transform
  coding of audio,'' in \emph{Audio Engineering Society Convention 141}, Sep.
  2016.

\bibitem{KankanahalliS2018icassp}
S.~Kankanahalli, ``End-to-end optimized speech coding with deep neural
  networks,'' in \emph{Proc. of the IEEE International Conference on Acoustics,
  Speech, and Signal Processing (ICASSP)}, 2018.

\bibitem{LiuQ2017perceptually}
Q.~J. Liu, W.~W. Wang, P.~J. Jackson, and Y.~Tang, ``A perceptually-weighted
  deep neural network for monaural speech enhancement in various background
  noise conditions,'' in \emph{2017 25th European Signal Processing Conference
  (EUSIPCO)}, 2017, pp. 1270--1274.

\bibitem{ZhenK2020spl}
K.~Zhen, M.~S. Lee, J.~Sung, S.~Beack, and M.~Kim, ``Psychoacoustic calibration
  of loss functions for efficient end-to-end neural audio coding,'' \emph{IEEE
  Signal Processing Letters}, vol.~27, pp. 2159--2163, 2020.

\bibitem{byun2022perceptual}
J.~Byun, S.~Shin, J.~Sung, S.~Beack, and Y.~Park, ``{Optimization of Deep
  Neural Network (DNN) Speech Coder Using a Multi Time Scale Perceptual Loss
  Function},'' in \emph{Proc. Interspeech 2022}, 2022, pp. 4411--4415.

\bibitem{defossez2022highfi}
A.~Défossez, J.~Copet, G.~Synnaeve, and Y.~Adi, ``High fidelity neural audio
  compression,'' \emph{arXiv preprint arXiv:2210.13438}, 2022.

\bibitem{hasegawa2003fundamentals}
M.~Hasegawa-Johnson and A.~Alwan, \emph{Speech Coding: Fundamentals and
  Applications}.\hskip 1em plus 0.5em minus 0.4em\relax John Wiley \& Sons,
  Ltd, 2003.

\bibitem{OordA2016wavenet}
A.~{van den Oord} \emph{et~al.}, ``{WaveNet: A Generative Model for Raw
  Audio},'' in \emph{Proc. 9th ISCA Workshop on Speech Synthesis Workshop (SSW
  9)}, 2016, p. 125.

\bibitem{KleijnW2018wavenet}
W.~B. Kleijn \emph{et~al.}, ``Wave{N}et based low rate speech coding,'' in
  \emph{Proc. of the IEEE International Conference on Acoustics, Speech, and
  Signal Processing (ICASSP)}, 2018, pp. 676--680.

\bibitem{GarbaceaC2019vqvae}
Y.~L. C.~Garbacea, A.~{van den Oord}, ``Low bit-rate speech coding with
  {VQ-VAE} and a wavenet decoder,'' in \emph{Proc. of the IEEE International
  Conference on Acoustics, Speech, and Signal Processing (ICASSP)}, 2019.

\bibitem{mp3}
{{ISO/IEC} 11172-3:1993}, ``{Coding of moving pictures and associated audio for
  digital storage media at up to about 1.5 Mbit/s},'' 1993.

\bibitem{HerreJ2008he-aac}
J.~Herre and M.~Dietz, ``Mpeg-4 high-efficiency aac coding [standards in a
  nutshell],'' \emph{IEEE Signal Processing Magazine}, vol.~25, no.~3, pp.
  137--142, 2008.

\bibitem{sbr}
{ISO/IEC 14496-3:2001/Amd 1:2003}, ``Information technology — coding of
  audio-visual objects — part 3: Audio — amendment 1: Bandwidth
  extension,'' 2003.

\bibitem{SunD2013mlsp}
D.~L. Sun and R.~Mazumder, ``Non-negative matrix completion for bandwidth
  extension: A convex optimization approach,'' in \emph{Proc. of the IEEE
  Workshop on Machine Learning for Signal Processing (MLSP)}, 2013.

\bibitem{EskimezS2019bandwidthex}
S.~E. {Eskimez}, K.~{Koishida}, and Z.~{Duan}, ``Adversarial training for
  speech super-resolution,'' \emph{IEEE Journal of Selected Topics in Signal
  Processing}, vol.~13, no.~2, pp. 347--358, May 2019.

\bibitem{gupta2019speechbe}
A.~Gupta, B.~Shillingford, Y.~Assael, and T.~C. Walters, ``Speech bandwidth
  extension with wavenet,'' in \emph{2019 IEEE Workshop on Applications of
  Signal Processing to Audio and Acoustics (WASPAA)}, 2019, pp. 205--208.

\bibitem{su2021bandwidth}
J.~Su, Y.~Wang, A.~Finkelstein, and Z.~Jin, ``Bandwidth extension is all you
  need,'' in \emph{ICASSP 2021 - 2021 IEEE International Conference on
  Acoustics, Speech and Signal Processing (ICASSP)}, 2021, pp. 696--700.

\bibitem{kim2022recovery}
J.-W. Kim, S.~K. Beack, W.~Lim, and H.~Park, ``Highly efficient audio coding
  with blind spectral recovery based on machine learning,'' \emph{IEEE Signal
  Processing Letters}, vol.~29, pp. 1212--1216, 2022.

\bibitem{shin2019recovery}
S.-H. Shin, S.~K. Beack, T.~Lee, and H.~Park, ``Audio coding based on spectral
  recovery by convolutional neural network,'' in \emph{ICASSP 2019 - 2019 IEEE
  International Conference on Acoustics, Speech and Signal Processing
  (ICASSP)}, 2019, pp. 725--729.

\bibitem{shin2020enhanced}
S.-H. Shin, S.~K. Beack, W.~Lim, and H.~Park, ``Enhanced method of audio coding
  using cnn-based spectral recovery with adaptive structure,'' in \emph{ICASSP
  2020 - 2020 IEEE International Conference on Acoustics, Speech and Signal
  Processing (ICASSP)}, 2020, pp. 351--355.

\bibitem{PetermannD2021harpnet}
D.~Petermann, S.~Beack, and M.~Kim, ``{HARP-Net}: Hyper-autoencoded
  reconstruction propagation for scalable neural audio coding,'' in \emph{Proc.
  of the IEEE Workshop on Applications of Signal Processing to Audio and
  Acoustics (WASPAA)}, 2021.

\bibitem{AgustssonE2017softmax}
E.~Agustsson \emph{et~al.}, ``Soft-to-hard vector quantization for end-to-end
  learning compressible representations,'' in \emph{Advances in Neural
  Information Processing Systems (NIPS)}, 2017, pp. 1141--1151.

\bibitem{RonnebergerO2015unet}
O.~Ronneberger, P.~Fischer, and T.~Brox, ``U-net: Convolutional networks for
  biomedical image segmentation,'' in \emph{International Conference on Medical
  image computing and computer-assisted intervention}.\hskip 1em plus 0.5em
  minus 0.4em\relax Springer, 2015, pp. 234--241.

\bibitem{pons2021upsampling}
J.~Pons, S.~Pascual, G.~Cengarle, and J.~Serrà, ``Upsampling artifacts in
  neural audio synthesis,'' in \emph{ICASSP 2021 - 2021 IEEE International
  Conference on Acoustics, Speech and Signal Processing (ICASSP)}, 2021, pp.
  3005--3009.

\bibitem{KingmaD2015adam}
D.~Kingma and J.~Ba, ``Adam: A method for stochastic optimization,'' in
  \emph{Proc. of the International Conference on Learning Representations
  (ICLR)}, 2015.

\bibitem{mushra}
{{ITU-R} {Recommendation} {BS} 1534-1}, ``Method for the subjective assessment
  of intermediate quality levels of coding systems ({MUSHRA}),'' 2003.

\end{thebibliography}
\end{document}